\documentclass[twocolumn,english,table,dvipsnames,svgnames]{aastex62}
\usepackage[T1]{fontenc}
\usepackage[latin9]{inputenc}
\usepackage{textcomp}
\usepackage{amssymb}
\usepackage{graphicx}
\usepackage{wasysym}

\makeatletter

\shorttitle{How common are SMBHs?}
\shortauthors{Buchner et al.}
\received{}
\revised{}
\accepted{}

\makeatother

\usepackage{babel}
\begin{document}
\selectlanguage{english}

\title{On the Prevalence of Super-Massive Black Holes over Cosmic Time}
\correspondingauthor{Johannes Buchner}
\email{johannes.buchner.acad@gmx.com}
\shorttitle{How common are SMBHs?}
\shortauthors{Buchner et al.}

\author[0000-0003-0426-6634]{Johannes Buchner}

\affiliation{Pontificia Universidad Católica de Chile, Instituto de Astrofísica,
Casilla 306, Santiago 22, Chile}

\author[0000-0001-7568-6412]{Ezequiel Treister}

\affiliation{Pontificia Universidad Católica de Chile, Instituto de Astrofísica,
Casilla 306, Santiago 22, Chile}

\author[0000-0002-8686-8737]{Franz E. Bauer}

\affiliation{Pontificia Universidad Católica de Chile, Instituto de Astrofísica,
Casilla 306, Santiago 22, Chile}

\affiliation{Millenium Institute of Astrophysics, Vicuña MacKenna 4860, 7820436
Macul, Santiago, Chile}

\affiliation{Space Science Institute, 4750 Walnut Street, Suite 205, Boulder,
Colorado 80301}

\author{Lia F. Sartori}

\affiliation{Institute for Particle Physics and Astrophysics, ETH Zürich, Wolfgang-Pauli-Str.
27, CH-8093 Zürich, Switzerland}

\author{Kevin Schawinski}

\affiliation{Institute for Particle Physics and Astrophysics, ETH Zürich, Wolfgang-Pauli-Str.
27, CH-8093 Zürich, Switzerland}
\begin{abstract}
We investigate the abundance of Super-Massive Black Hole (SMBH) seeds
in primordial galaxy halos. We explore the assumption that dark matter
halos outgrowing a critical halo mass $M_{c}$ have some probability
$p$ of having spawned a SMBH seed. Current observations of local,
intermediate-mass galaxies constrain these parameters: For $M_{c}=10^{11}M_{\odot}$,
all halos must be seeded, but when adopting smaller $M_{c}$ masses
the seeding can be much less efficient. The constraints also put lower
limits on the number density of black holes in the local and high-redshift
Universe. Reproducing $z\sim6$ quasar space densities depends on
their typical halo mass, which can be constrained by counting nearby
Lyman Break Galaxies and Lyman Alpha Emitters. For both observables,
our simulations demonstrate that single-field predictions are too
diverse to make definitive statements, in agreement with mixed claims
in the literature. If quasars are not limited to the most massive
host halos, they may represent a tiny fraction ($\approx10^{-5}$)
of the SMBH population. Finally, we produce a wide range of predictions
for gravitational events from SMBH mergers. We define a new diagnostic
diagram for LISA to measure both SMBH space density and the typical
delay between halo merger and black hole merger. While previous works
have explored specific scenarios, our results hold independent of
the seed mechanism, seed mass, obscuration, fueling methods and duty
cycle.
\end{abstract}

\keywords{quasars: supermassive black holes, galaxies: halos, galaxies: high-redshift,
galaxies: evolution}

\date{-Received date / Accepted date}

\section{Introduction}

Supermassive Black Holes (SMBHs) are ubiquitous in local, massive
galaxies \citep{Kormendy2013,Graham2015}. This raises the question
of how these black holes came into existance, and how they evolved
over cosmic time. Mass growth of black holes over cosmic time is thought
to be traced by active galactic nuclei \citep[AGN; e.g.,][]{Soltan1982,YuTremaine2002}.
The physical process creating black holes in the first place remains
unknown, although multiple scenarios have been proposed, including
collapses in the cores of massive stars, dense nuclear star clusters
or pristine gas clouds \citep{Rees1984,Latif2016}. These lead to
different initial mass regimes. Much research has been dedicated to
develop detailed physically meaningful seed mechanisms and timely
mass growth \citep[see e.g.,][]{Volonteri2010,Reines2016}. This work
however addresses a simpler problem: How frequently do massive black
holes need to arise to explain current observations?

Modern cosmological simulations now routinely incorporate SMBHs to
quench star formation in massive halos \citep[see e.g.][]{Croton2006}.
For computational convenience, the most common method of creating
SMBHs is to seed dark matter halos once they reach a certain critical
mass, $M_{c}$, e.g., $M_{h}>M_{c}=10^{10-11}M_{\odot}$ \citep{Somerville2015}.
However, massive galaxies in the local Universe have been assembled
from many such progenitor halos. \citet{Menou2001} computed that
one can achieve high SMBH occupation fractions at the massive end
by only seeding 3\% of $M_{c}\approx10^{9}M_{\odot}$ halos at redshift
$z=5$. They demonstrated this by letting the seeds trickle down simulated
merger trees, the representation of the hierarchical galaxy evolution.
They then made predictions about the rate of SMBH births over cosmic
time, merger rates (relevant for space-based long-baseline gravitational
wave experiments). Since then, our understanding of the cosmology
has improved, numerical simulations have become more sophisticated
and local observations give a better idea down to which mass limit
SMBHs occupy galaxies.

While dedicated simulations tend to be highly specific to a particular
seeding and feeding scenario \citep[see, e.g.,][]{Volonteri2010,Naab2017},
occupation calculations under this framework are independent of the
mass of the seeding process and the mass growth of these seeds. All
that matters is whether a massive black hole is present or not.

Based on reliable halo trees from dark-matter simulations presented
in §\ref{sec:sims}, this work makes three main contributions: 1)
Section \ref{sec:dwarfs} explores how efficient seeding mechanisms
need to be to explain current black hole occupation observations.
2) Section \ref{sec:evol} explores the evolution of the required
black hole population. In §\ref{sec:gw} we make predictions for the
future gravitational wave experiment LISA, and demonstrate how the
observed events can be used to infer both the space density of SMBHs
and merger inspiral delays. 3) Section \ref{sec:z6qsos} focuses on
the environment of high-redshift quasars, and clarifies how nearby
galaxies trace the halo mass of quasar hosts.

\section{Simulations}

\label{sec:sims}

The premise of our analysis is that galaxy halos that grow to a critical
mass $M_{c}$ (e.g., $10^{10}M_{\odot}$) have a chance probability
$p$ to have produced a seed black hole by this point. For an individual
galaxy the likelihood of forming a seed black hole will in reality
depend in some unknown manner on other properties (e.g.~environment,
radiation fields, gas metallicity, evolutionary history). Considering
however the halo population crossing a given mass threshold, we can
define an effective occurance rate $p$ up to that evolutionary stage.
If the future evolution of the halo and the seeding probability correlate
only weakly, this formalism allows an estimate of the efficiency of
the seeding process. We explore the implications of such seeding recipes
and viable parameter ranges for $M_{c}$ and $p$.

\begin{figure}
\begin{centering}
\includegraphics[width=1\columnwidth]{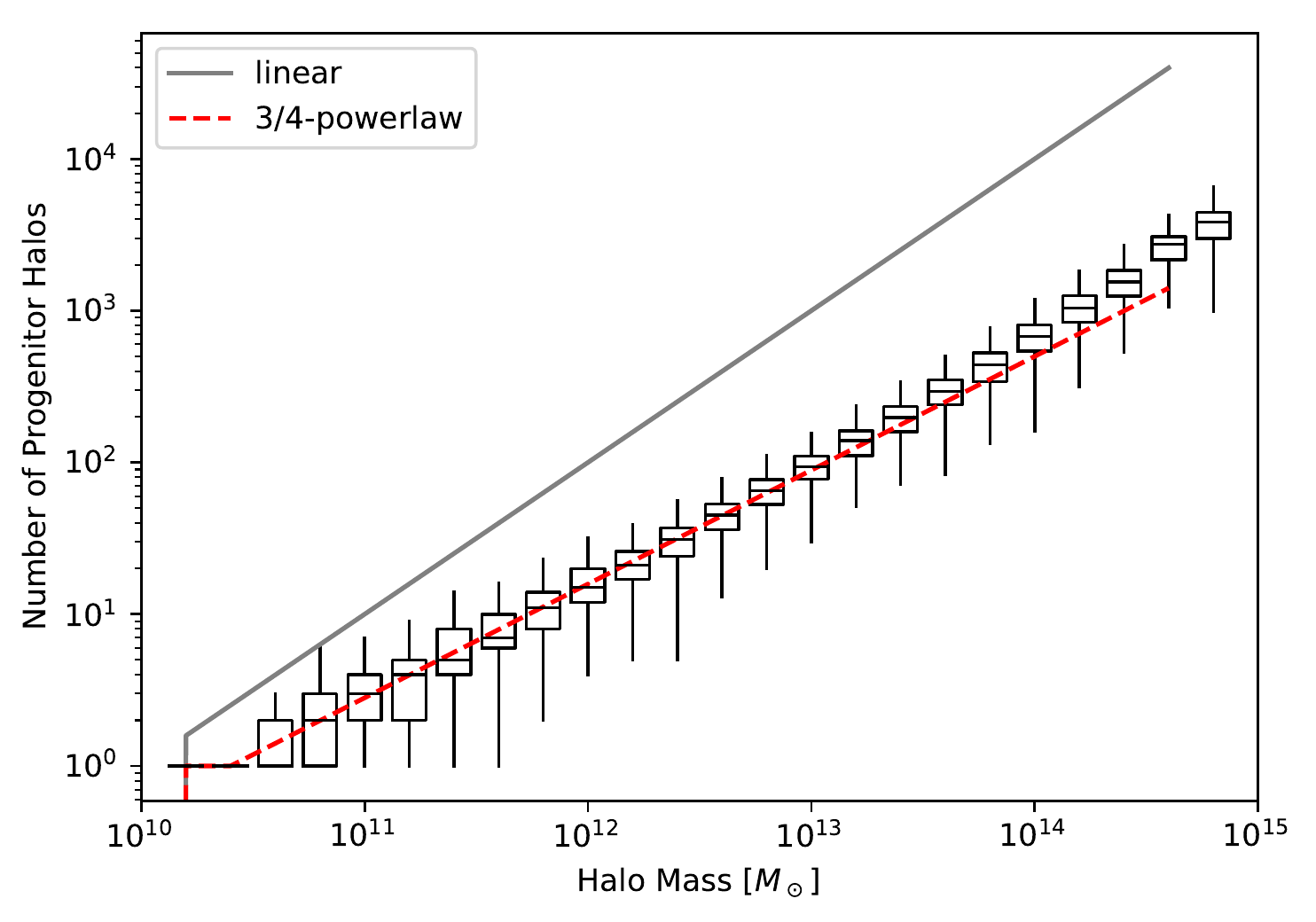}
\par\end{centering}
\caption{\label{fig:nhalos}Number of $10^{10}M_{\odot}$ halos that built
a $z=0$ halo. At each halo mass bin, we count the number of progenitors
that merged into each halo. The rectangles with horizontal line represent
the 25\%-75\% quantile range and median of each mass bin. The vertical
lines shows the range of the distribution. The red dashed line indicates
a 3/4 powerlaw relation.}
\end{figure}

Connecting (potentially early) black hole seeding with observations
in the local universe requires simulations that follow the mass evolution
of the Universe. High mass resolution is necessary to follow sites
of emerging proto-galaxies. Additionally, because seeds may be rare,
large cosmological volumes need to be probed. To this end, we use
the \emph{MultiDark} cosmological dark matter N-body simulations \citep{Prada2012,Klypin2016}.
These assume a \emph{Planck} cosmology ($h=0.6777$, $\Omega_{\Lambda}=0.693$,
$\Omega_{m}=0.307$, \citealp{PlanckCollaboration2014}), which we
use throughout\footnote{We explicitly multiply out $h$ e.g. in distances, but for comparison
with literature we keep halo masses in units of $h^{-1}$.}. The simulation evolves an initial dark matter density distribution
over cosmic time under gravity. This encompasses the gravity-dominated
collapse into sheets, filaments and finally halos, wherein galaxies
should reside, as well as the mergers of structures. To study the
evolution of halos, merger trees\footnote{Constructed with \emph{ROCKSTAR} and \emph{ConsistentTrees} \citep{Behroozi2013,Behroozi2013a}.}
summarise the merging of (sub-)halos as well as the dark matter halo
mass at each simulation snapshot. In particular we focus on the highest-resolution,
Small MultiDark simulation (SMDPL) with a box size of $400\mathrm{Mpc/h}$,
populated with $3840^{3}$ dark matter particles of mass $10^{8}M_{\odot}/h$,
which resolves well halos of masses down to $10^{10}M_{\odot}$. When
smaller halo resolution is needed, we use the $40\mathrm{Mpc}/h$
``Following ORbits of Satellites'' \citep[FORS;][]{Gonzalez2016}
simulation, whose small particle masses ($\approx4\times10^{6}M_{\odot}/h$)
resolve halos down to $10^{8}M_{\odot}$ well. The FORS cosmology
is very similar to the above-mentioned values. This work adopts virial
masses in $M_{\odot}h^{-1}$ throughout\footnote{To convert into physical units, our reported masses need to be multiplied
by $h$ and divided by $\sim0.75$ to account for the impact of baryons
\citep[see e.g.,][]{Sawala2013}.} and physical units for all other quantities. 

Throughout, we refer to the created black holes as seeds or (S)MBHs.
However, because we only count the halo occupation, our model does
not need to assign any specific mass to them. Our calculations are
thus \emph{independent of seed mass} and mass evolution. Consequently,
we refrain from exploring correlations of galaxy properties with black
hole mass. By-products of the seeding process that do not become MBHs
are not considered here, but may additionally be present in the Universe.

\section{SMBHs in local galaxies}

\label{sec:dwarfs}

\begin{figure}
\begin{centering}
\includegraphics[width=1\columnwidth]{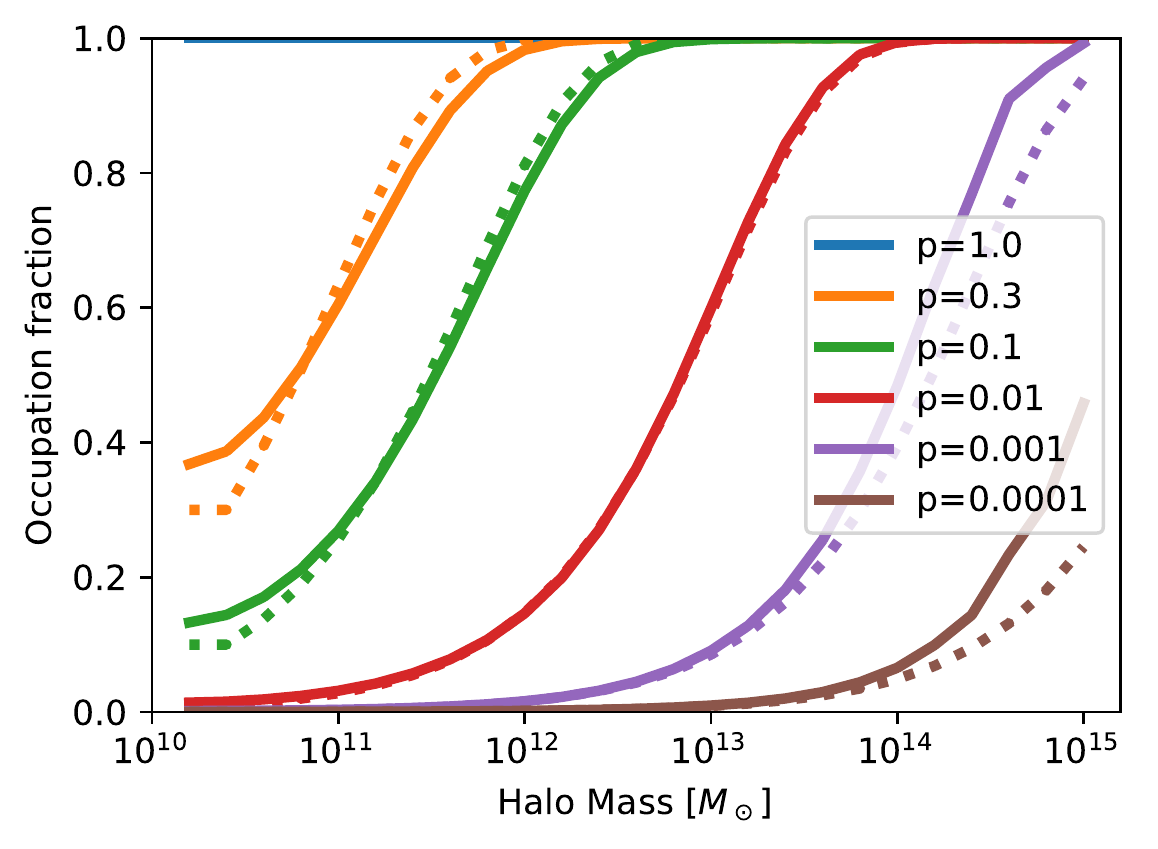}
\par\end{centering}
\caption{\label{fig:frachalos0-1}Black hole occupation in the local Universe
as function of halo mass. The solid curves are from dark matter simulations.
The dotted curves are computed from the analytic equations~\ref{eq:simpleanalytic}~and~\ref{eq:occfrac}.
This figure assumes $M_{c}=10^{10}M_{\odot}$.}
\end{figure}

\begin{figure*}
\begin{centering}
\includegraphics[width=0.96\textwidth]{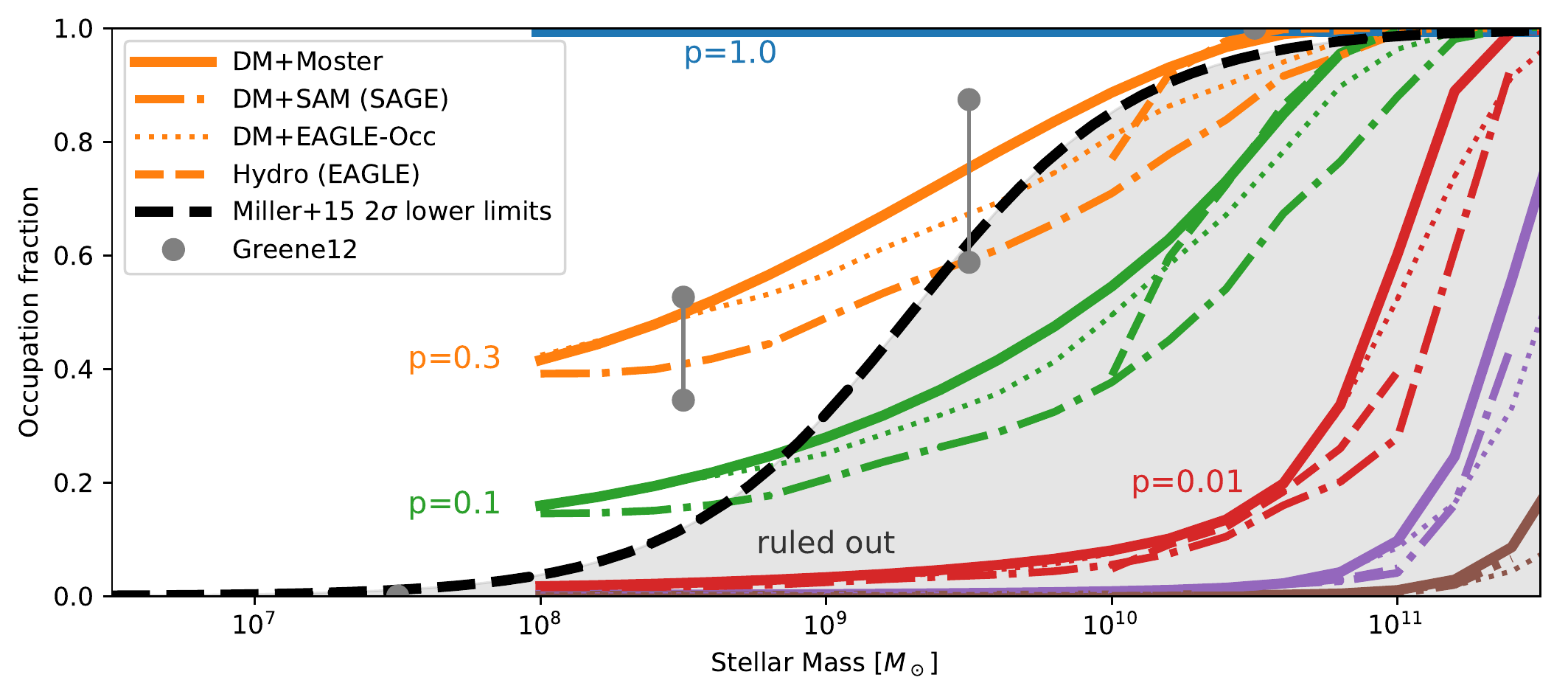}
\par\end{centering}
\caption{\label{fig:frachalos0}Black hole occupation in the local Universe
as function of stellar mass. Differently coloured model curves correspond
to different seeding probabilities $p$. Observational lower limits
are shown as a dashed black curve \citep[$2\sigma$ lower limits]{Miller2015}
and gray connected points \citep{Greene2012}. These imply $p\gtrsim30\%$
for the chosen mass limit ($M_{c}=10^{10}M_{\odot}$). Figure~\ref{fig:fracparamspace}
explores other mass limits.}
\end{figure*}

\begin{figure}
\begin{centering}
\includegraphics[width=1.05\columnwidth]{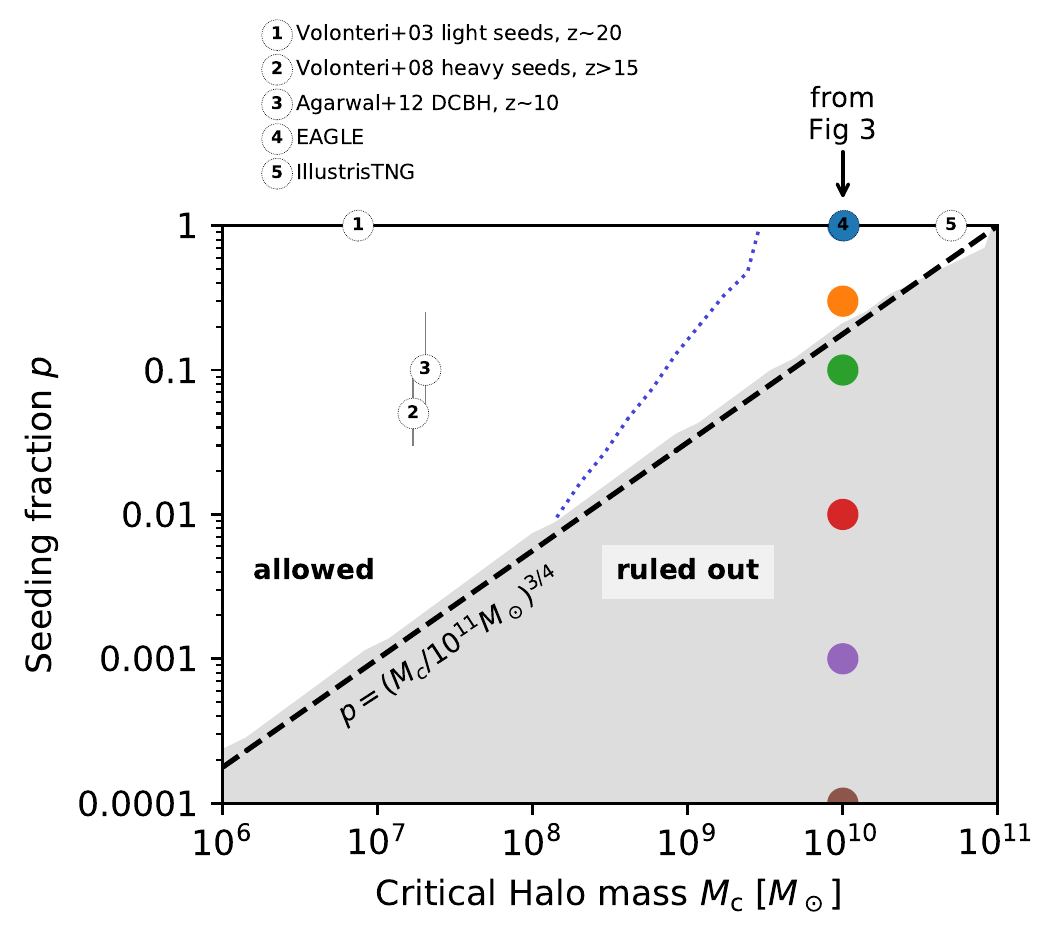}
\par\end{centering}
\caption{\label{fig:fracparamspace}Allowed parameter region (white area) given
the M15 limits (see Figure~\ref{fig:frachalos0}). The colored dots
are the parameters shown in Figure~\ref{fig:frachalos0} as thick
lines, with the same colors. If seeding is only allowed at $z>7$,
the allowed region shrinks, as indicated by the blue dotted line.
Circled numbers allow approximate comparison to some physical seeding
models \citep{Volonteri2003,Volonteri2008,Agarwal2014} and large-scale
cosmological simulations \citep{Weinberger2018,Schaye2015,Crain2015}.}
\end{figure}

Owing to the hierarchical growth of structures, a high number of halos
go into building massive galaxies. When halos merge, occupant black
holes are generally passed on to the merger product (this assumption
is discussed below). This implies that the fraction of halos that
need to be seeded could be very low \citep{Menou2001}. As a starting
point, we simply count for $z=0$ halos, from how many $10^{10}M_{\odot}$
halos they were built. Specifically, we count in each merger tree
the leaves with $M>M_{c}$ . This is shown in Figure~\ref{fig:nhalos}.
If galaxies only were built by mergers of existing halos, we would
expect a linear relation, $N_{z=0}=M_{z=0}/M_{c}$ (gray line). However,
growth through accretion of non-collapsed structures (e.g., filaments)
is also important. A better empirical description is a sub-linear
powerlaw relation with sub-unity normalisation:
\begin{equation}
N_{z=0}=\frac{1}{2}(M_{z=0}/M_{c})^{3/4}\label{eq:simpleanalytic}
\end{equation}
This is plotted as a red dashed curve in Figure~\ref{fig:nhalos},
and is an appropriate approximation as we will show below. Around
this mean number of constituents, Figure~\ref{fig:nhalos} shows
modest scatter of approximately $0.2\mathrm{dex}$. We verified that
this relation also holds for other values of $M_{c}$ ($10^{8.5-11}M_{\odot}$).

We then compute the black hole occupation fraction. If a halo is made
from $N$ building blocks, each with equal chance $p$ to contain
a seed black hole, it contains a black hole with probability:
\begin{equation}
P(\mathrm{BH}|N)=1-\left(1-p\right)^{N}\label{eq:occfrac}
\end{equation}
This is derived by considering the probability that none of the $N$
halos have a MBH, and taking the complement. A subtle point here is
that we do not need to keep track of the number of black holes inside
a halo and whether/when they merge or get ejected from the system.
Equation~\ref{eq:occfrac} only assumes that if multiple halos with
black holes merge, \emph{at least one} of these black holes remains
in the merged halo. The average of $P(BH|N)$ over a halo population
yields the expected occupation fraction. Figure~\ref{fig:frachalos0-1}
plots the occupation fraction as a function of halo mass and seeding
fraction $p$, for the case of $M_{c}=10^{10}M_{\odot}$. We now constrain
$p$ using observations.

We focus on a recent measurement of the fraction of local galaxies
containing a massive black hole as a function of mass by \citet[M15]{Miller2015}.
They surveyed local early-type galaxies for central X-ray point sources,
a telltale sign of accretion onto SMBHs, and carefully corrected for
flux limits. Correcting for the X-ray inactive fraction of black holes
is systematically uncertain. The correction anchors on the X-ray active
fraction at the higher masses, where the occupation fraction is thought
to be $\sim100\%$ \citep{Greene2012}. M15 used an advanced methodology
to incorporate that the X-ray luminosity distribution  changes with
host mass. They detect sources down to and below host galaxy stellar
masses of $10^{9}M_{\odot}$ and their inferred lower limits on the
black hole occupation fraction are shown in the bottom panel of Figure~\ref{fig:frachalos0}
as a thick black dashed line. \citet{Greene2012} previously reviewed
observational constraints and also considered the late-type spiral
sample of \citet{Desroches2009}, obtaining slightly elevated lower
limits (gray connected points in Figure~\ref{fig:frachalos0}).

To compare our simulation to these observational constraints, we convert
from halo masses to stellar masses. We test several methods: Firstly,
we assign stellar masses from the \citet{Moster2010} conditional
stellar mass function $P(M_{\star}|M_{h})$\footnote{We only consider the central galaxies because we work with a subhalo
catalogue and satellites are unimportant in the mass range of interest.} derived from matching the local stellar mass function and galaxy
clustering to simulated dark matter halos. Importantly, this empirical
method does not assume any galaxy evolution physics and is observational.
Secondly, we try assigning stellar masses according to the distribution
produced by the SAGE semi-analytic model for the MDPL2 simulation
\citep{Croton2016} and that produced by the EAGLE hydro-radiative
simulation. The latter \citep[version Recal-L025N0752]{Schaye2015}
reproduces the galaxy mass function and sizes very well, in particular
in this mass regime. However, populating halos based only on a $M_{\star}-M_{h}$
distribution neglects that the number of progenitors $N$ could influence
$M_{\star}$ at a given $M_{h}$, inducing a correlation between occupation
probability and $M_{\star}$. To test this, we also derive the number
of progenitors~$N$ and $M_{\star}$ from the merger trees of EAGLE.
The groups of curves in Figure~\ref{fig:frachalos0} demonstrate
that for a given $p$ the results from these four methods are consistent.
Thus, the choice of $M_{h}-M_{\star}$ conversion method does not
materially affect our conclusions.

\begin{figure*}
\begin{centering}
\includegraphics[width=0.334\textwidth]{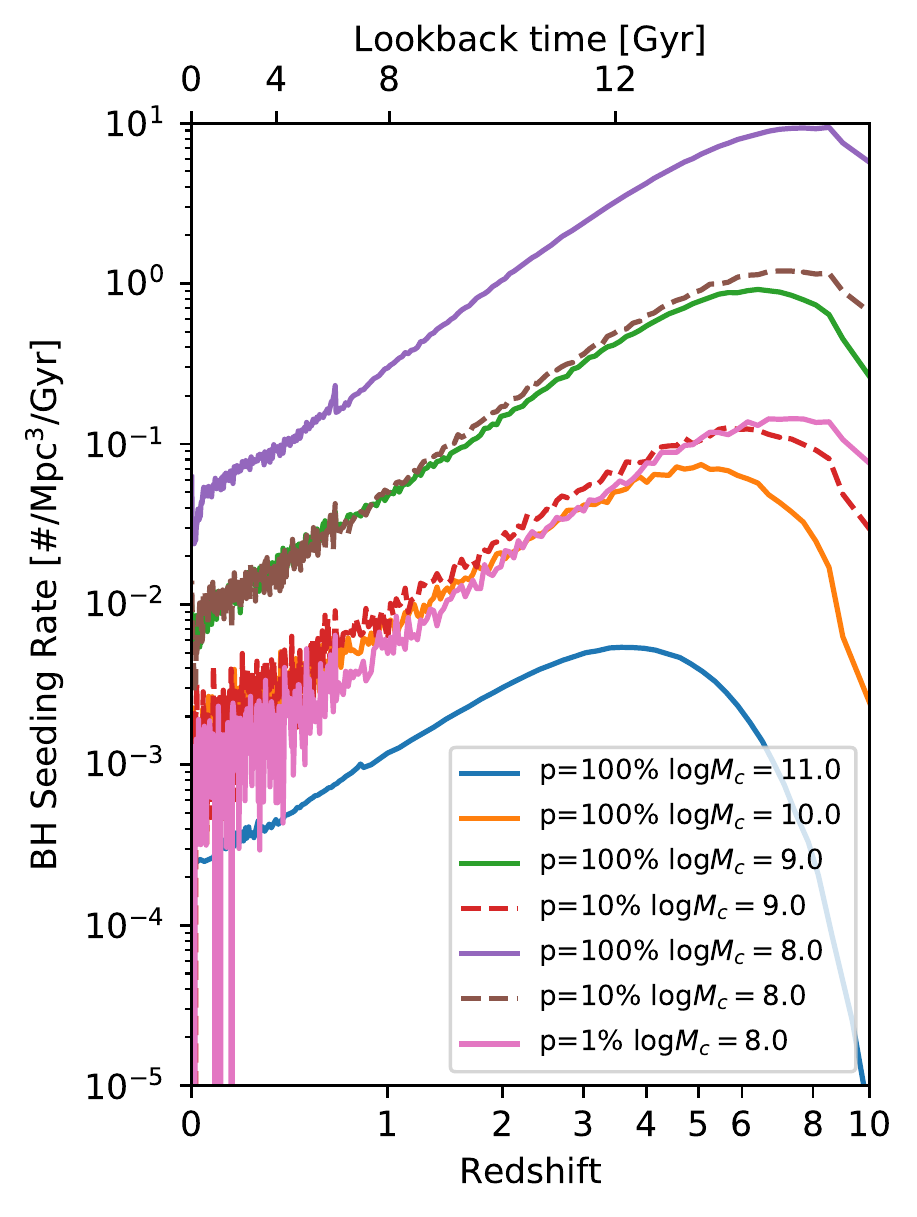}\includegraphics[width=0.334\textwidth]{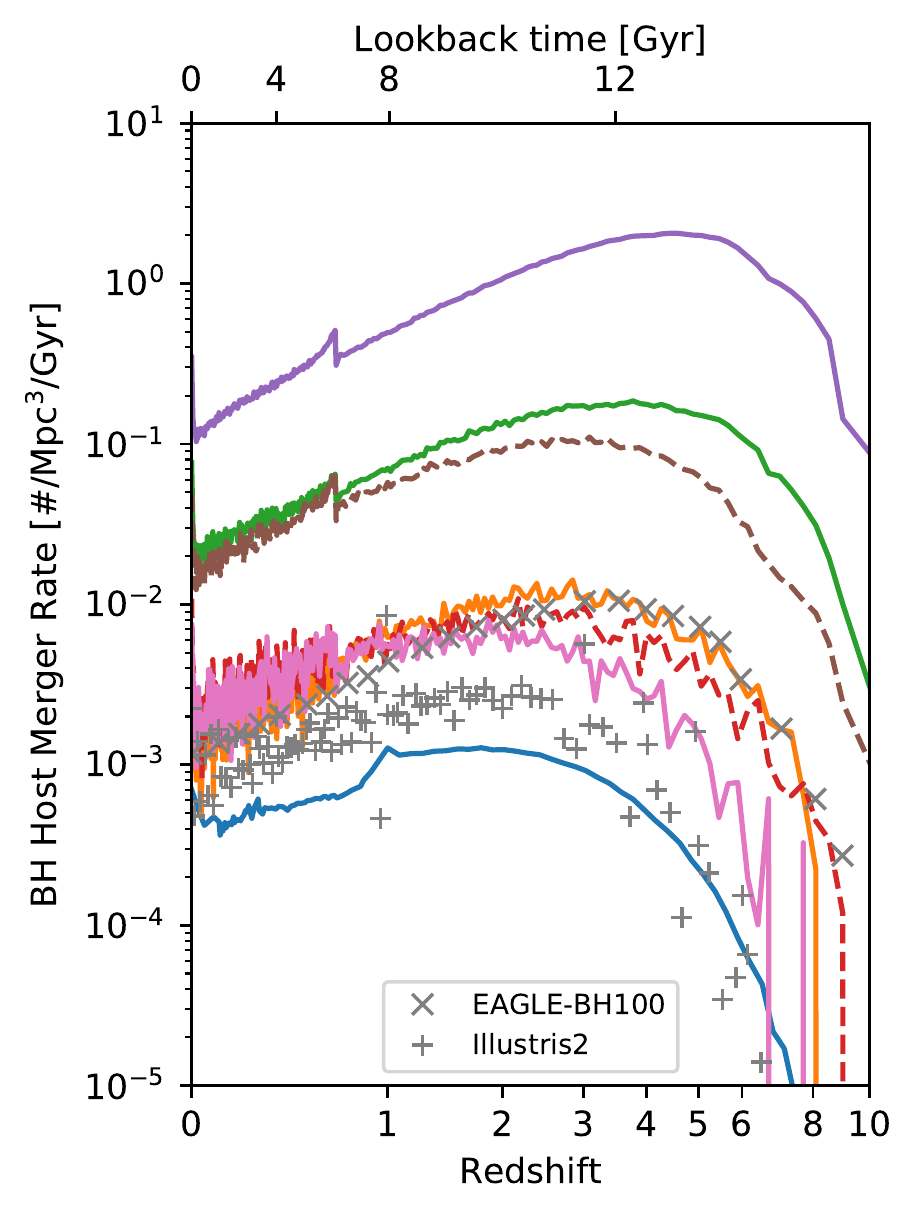}\includegraphics[width=0.334\textwidth]{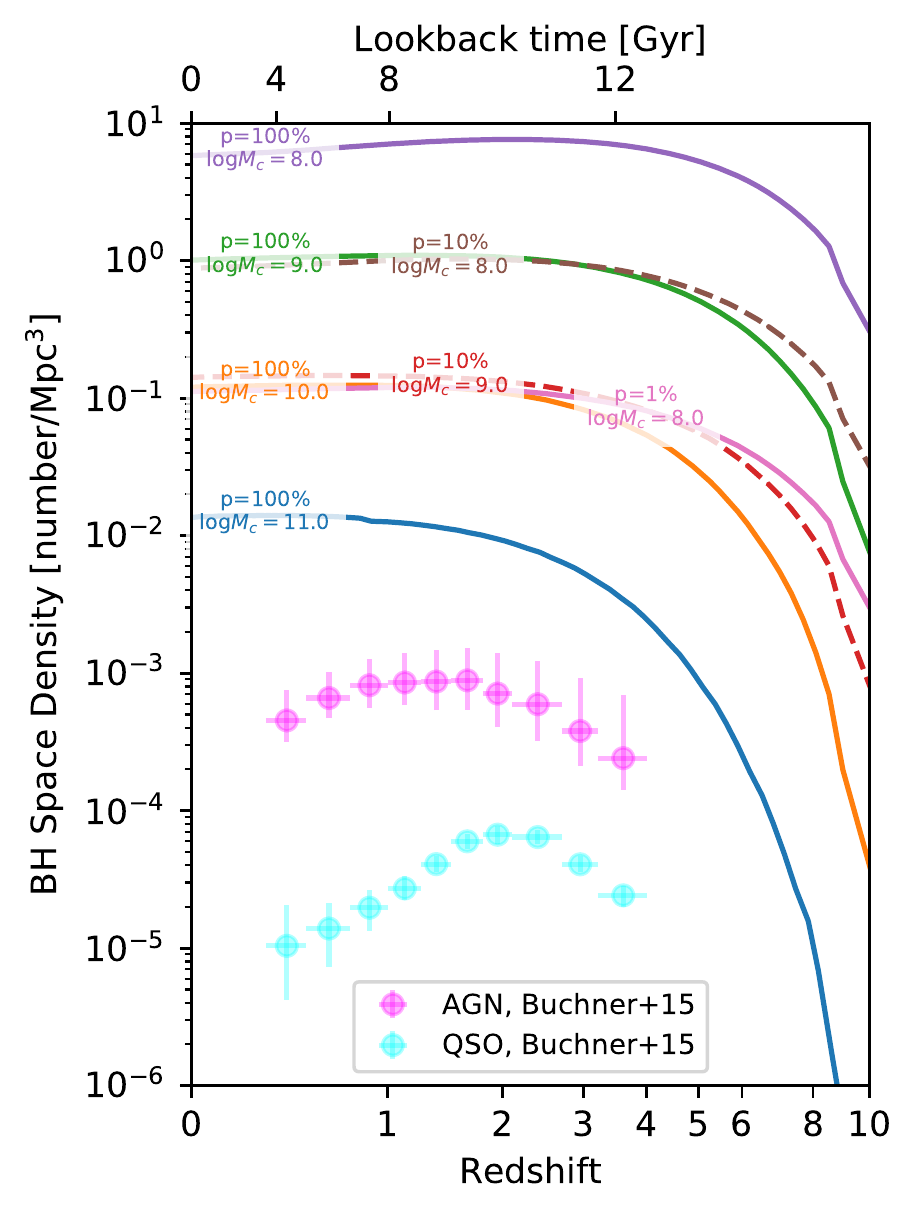}
\par\end{centering}
\caption{\label{fig:evolution}\emph{Left panel}: The birth rate of black holes
over cosmic time. Curves show different seeding fractions $p$ and
mass thresholds $M_{0}$. \emph{Center panel}: Merger rate of halos
occupied by black holes. The kink at $z\sim1$ is due to the simulations
cosmic variance, i.e., a large overdensity merging at that redshift.
\emph{Right panel}: The total number of halos with black holes. We
compare with space-densities of AGN with luminosities $L_{\mathrm{X}}>3\times10^{42}\mathrm{erg/s}$
(AGN) and $L_{\mathrm{X}}>10^{44}\mathrm{erg/s}$ (QSOs) measured
by X-ray surveys sensitive to unobscured and obscured accretion \citep{Buchner2015}.
Comparing the model curves and observations, we see that only a very
small fraction (1:10 to 1:10,000) of black holes accrete at these
luminosities.}

\vspace{0.6cm}
\end{figure*}

Figure~\ref{fig:frachalos0} shows the occupation fraction predictions
for $M_{c}=10^{10}M_{\odot}$. The comparison with the observational
lower limit implies $p\gtrsim30\%$, with lower seed probabilities
ruled out. For higher masses, even higher seed probabilities are necessary,
e.g., $M_{c}=10^{11}M_{\odot}$ matches observations only with full
occupation ($p=100\%$). We explore the allowed seeding fractions
$p$ by extrapolating equations~\ref{eq:simpleanalytic}~and~\ref{eq:occfrac}
into lower mass regimes. We compute curves like in Figure~\ref{fig:frachalos0}
with the mean $M_{\star}/M_{h}$ ratio of \citet{Moster2010} and
accept those that pass the M15 limits. Figure~\ref{fig:fracparamspace}
shows the allowed parameter space. Because the number of halo building
blocks has a $3/4$ powerlaw exponent (see Figure~\ref{fig:nhalos}
and eq.~\ref{eq:simpleanalytic}), we find the constraint:
\begin{equation}
p\apprge(M_{c}/10^{11}M_{\odot})^{3/4}.\label{eq:paramconstraint}
\end{equation}
In the following, we consider only models satisfying this constraint.

\section{The SMBH population over cosmic time}

\label{sec:evol}

To explore how the SMBH population evolves over cosmic time, we populate
the halo tree according to our recipe. When a halo for the first time
reaches a mass $M_{c}$, with chance $p$ it is marked as occupied
by a black hole. The occupation is inherited by the merger tree descendant.
As a consequence, in our seeding framework black holes are not only
``born'' in the high-redshift Universe. To show this, the left panel
of Figure~\ref{fig:evolution} presents birth rates over cosmic time.
Seeding is most frequent at early times ($z>4$). Depending on the
model parameters, the median seeding time lies in the second, third
or forth billion years. At $z<6$, the model curve shapes are almost
independent of the input parameters. Regarding the normalisation,
we note that a $100\%$ seeding above mass $M_{c}$ creates a black
hole population at a similar rate as $10\%$ seeding at mass $10\%\times M_{c}$
(left panel of Figure~\ref{fig:evolution}).

As halos containing black holes merge, interactions of multiple SMBHs
are possible. The middle panel of Figure~\ref{fig:evolution} shows
mergers of halos hosting black holes. The normalisation again scales
as just described for the birth rate. Changes in the critical mass
shift the peak slightly. We explore the possible merging of the black
holes themselves in the next section.

The right panel of Figure~\ref{fig:evolution} presents the total
number of massive black holes over cosmic time. For $p=1$, this is
just the number of halos above the mass threshold. Overall, we find
a present-day SMBH space density of $>0.01\,\mathrm{Mpc^{-3}}$ (with
models not ruled out in Figure~\ref{fig:fracparamspace}). In the
local Universe, the fraction of occupied halos is mass-dependent in
the way presented in Figure~\ref{fig:frachalos0}. The comparison
between the number of SMBHs to the observed number of AGN provides
the fraction of actively accreting black holes in excess of that luminosity
threshold. Modern surveys detect hard X-ray emission even in heavily
obscured AGN, and recover the intrinsic accretion luminosity using
X-ray spectra \citep[e.g.,][]{Buchner2015}. In Figure~\ref{fig:evolution}
we include the space density of AGN at $L(2-10\mathrm{keV})>3\times10^{42}\mathrm{erg/s}$
and $>10^{44}\mathrm{erg/s}$, which correspond approximately to accretion
rates of $5\times10^{6}$ and $5\times10^{8}M_{\odot}/\mathrm{Gyr}$
\citep{Marconi2004}, respectively, assuming 10\% radiative efficiency.
The AGN density inferred from observations is $1-3$ orders of magnitude
below the total SMBH population in Figure~\ref{fig:evolution}. The
comparison indicates that the vast majority of the SMBH population
is dormant or accretes at low levels.

\newpage{}

\section{Gravitational Waves from SMBH mergers}

\label{sec:gw}The merging of halos in principle brings their SMBHs
together as well. The merging of SMBHs produces gravitational waves,
detectable with the proposed Laser Interferometer Space Antenna \citep[LISA,][]{Amaro-Seoane2017}.
\citet{Salcido2016} predicted in detail waveforms of SMBH mergers
from the EAGLE cosmological simulation. Because of the sensitivity
down to $10^{4}M_{\odot}$ \citep{Amaro-Seoane2017}, \citet{Salcido2016}
find that essentially all mergers are within the sensitivity of LISA,
with their first mergers of the low-mass SMBH seeds dominating. Therefore,
one can neglect mass and sensitivity considerations and focus on the
occurance of SMBH merger events. We predict the number and redshifts
of observable GW events from our scenarios. In this computation, the
rest-frame black hole mergers at a given redshift $r(z)$ within a
redshift interval $dz$ in a comoving volume $dV_{c}$ are converted
into observable rates across the entire sky using $\frac{dz}{dt}\times\frac{dV_{c}}{dz}\times\frac{1}{1+z}$
to account for the corresponding rest-frame time, comoving volume
on the sky and time dilation \citep[e.g.,][]{Sesana2004,Salcido2016}. 

\begin{figure}
\begin{centering}
\includegraphics[width=1\columnwidth]{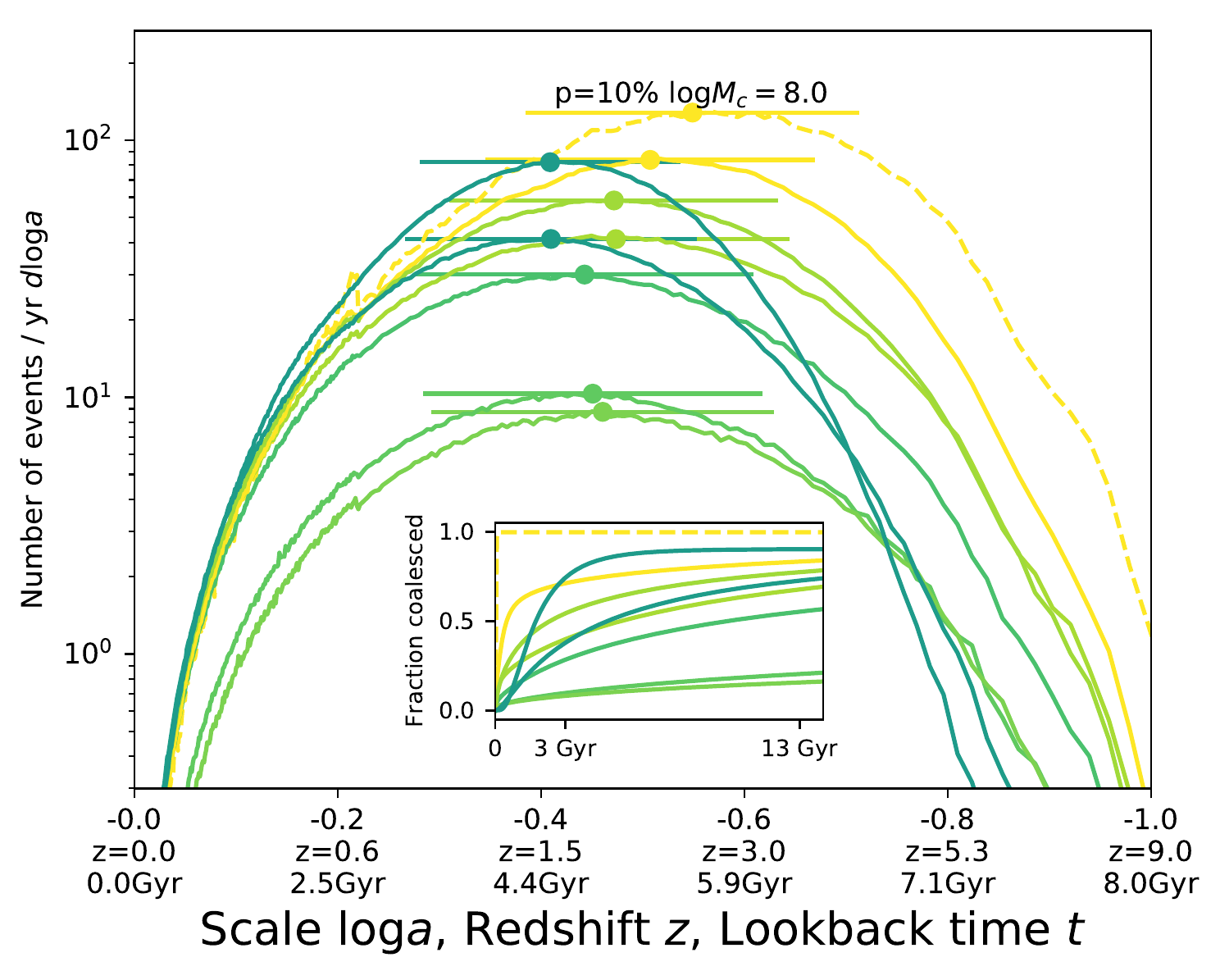}
\par\end{centering}
\caption{\label{fig:GW-z-distribution}Distribution of predicted Gravitational
Wave events in $\log(1+z)$. The 10\% seeding of $M_{0}=10^{8}M_{\odot}$
halos is shown here as the yellow dashed line. Different delay distributions
\citep[inset, from][]{Kelley2017} shift the distribution to later
times, and can reduce the number of events. The error bars show the
mean~$\mu$ and standard deviation $\sigma$ of the $\log(1+z)$
distributions, which are combined in Figure~\ref{fig:GW} as $\bar{A}=\mu-4\sigma$.
The colors indicate the location of the mean. Altering the seeding
prescription mostly changes the normalisation (see middle panel of
Figure~\ref{fig:evolution}).}
\end{figure}

\begin{figure*}
\begin{centering}
\includegraphics[width=0.75\textwidth]{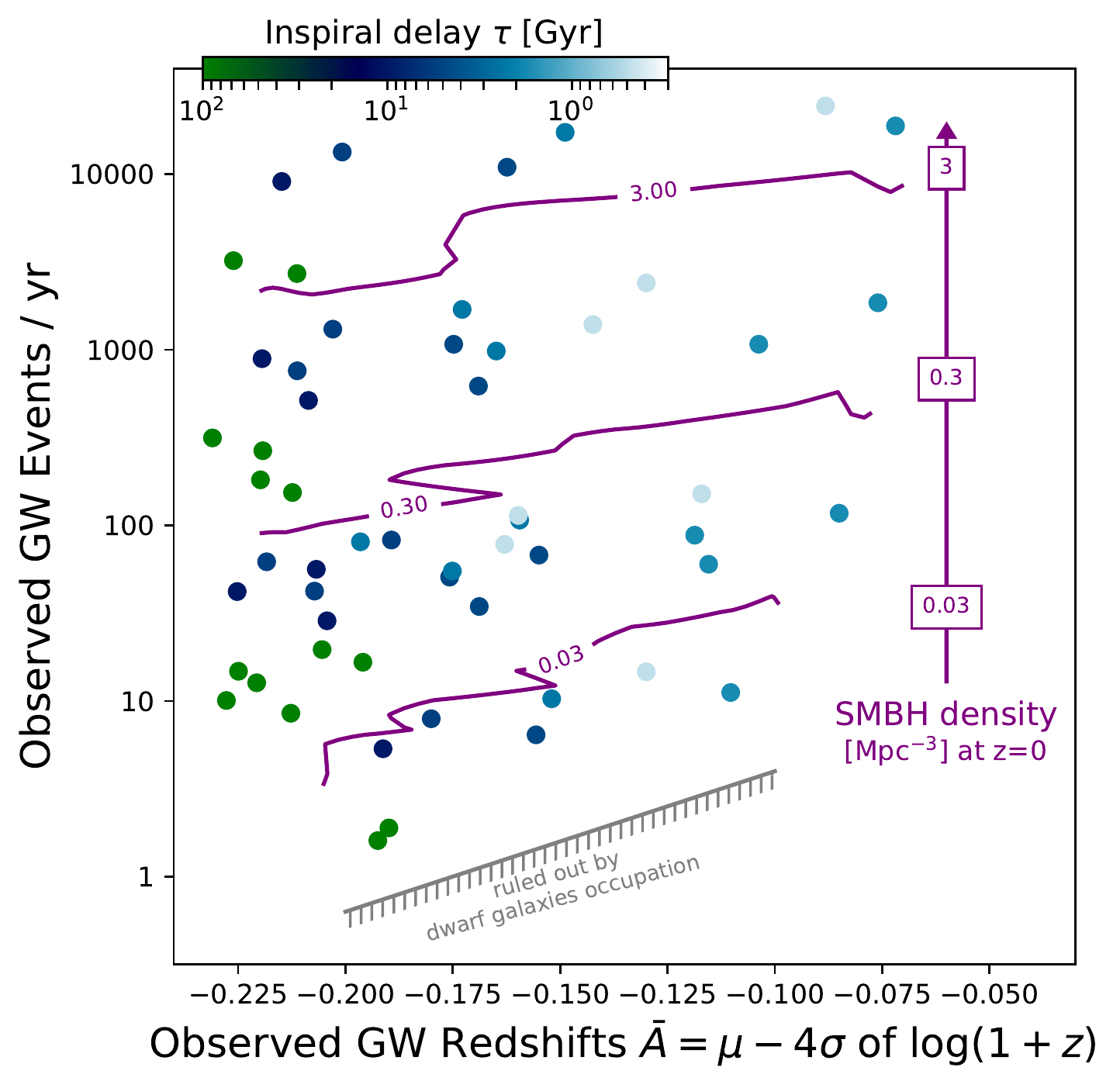}
\par\end{centering}
\caption{\label{fig:GW}Gravitational wave diagnostic diagram. The plot axes
are observables: The number of GW events per year (y-axis), and a
skew statistic of their redshift distribution (x-axis). Points correspond
to various seeding models and inspiral delay functions (color-coded
by median delay, from Figure~\ref{fig:GW-z-distribution}). Purple
contour curves indicate the number of black holes predicted by each
model, which influences strongly the GW event rate (y-axis), with
limited effect by the delay time. The observable statistic on the
x-axis approximately corresponds to the inspiral delay. Models below
1/yr are excluded by current occupation constraints (see Figure~\ref{eq:occfrac}).}
\end{figure*}
However, the time delay between halo merger and black hole merger
is uncertain. \citet{Kelley2017} explored in detail the physical
processes as SMBH binaries overcome nine orders of magnitude in separation
from kilo- to micro-parsecs. They also present several physically
reasonable \emph{delay functions}, quantifying the fraction of SMBH
binaries have merged after a given time. These are shown in the inset
of Figure~\ref{fig:GW-z-distribution}. These delay functions have
the effect of changing the observed merger redshift distribution in
a characteristic fashion. For example, the dashed curve in the main
panel of Figure~\ref{fig:GW-z-distribution} is assuming no delay
\citep[see also][]{Salcido2016} between halo merger and black hole
merger, while each solid curve applies a different delay function.
These generally move the peak to lower redshifts, narrow the distribution
and can suppress the rate of black hole mergers substantially. Overall,
all of the distributions appear approximately Gaussian in the scale
factor, $\log a=-\log(1+z)$.

We define two LISA observables: The number of GW events observable
per year, $N$, and, $\bar{A}$, which describes the location and
shape of the distribution and is defined as:
\begin{equation}
\bar{A}:=\mu-4\sigma.\label{eq:Abar}
\end{equation}
The $\bar{A}$ statistic combines the mean $\mu$ and standard deviation
$\sigma$ of the $\log(1+z)$ distribution (see for example Figure~\ref{fig:GW-z-distribution}).
Through experimentation, we found that an $\bar{A}$ centered on $\mu$
but skewed 4$\sigma$ to the left of the distribution captures well
both the shift and narrowing of the redshift distribution caused by
merger delays. It can be readily computed from detected gravitational
wave events, assuming a cosmology.

These two observables diagnose the underlying SMBH population and
differentiate delay functions. Figure~\ref{fig:GW} presents our
\emph{LISA diagnostic diagram}, populated by all combinations of the
seeding prescriptions and delay functions. The number of events predicted
(\emph{y-axis}) generally reflects the overall space density of SMBHs
(purple contour curves and arrow), and is thus primarily driven by
the chosen seeding prescription. It is notable that a wide range of
predictions are possible, from few to thousands of events per year.
However, even in the most unfavorable scenario with slow inspirals
and a high mass threshold ($10^{11}M_{\odot}$), a few detections
per year are predicted. The color-coding of the LISA model points
in Figure~\ref{fig:GW} indicates the median delay time corresponding
to a delay function in the inset of Figure~\ref{fig:GW-z-distribution}.
Because the colors approximately change along the x-axis, the $\bar{A}$
statistic (\emph{x-axis}) is a proxy for the delay function, and can
distinguish slow from quick inspirals (green to light blue points,
from left to right). To quantify $\bar{A}$ with an uncertainty of
$\pm0.05$ requires approximately 200 GW detections.

We verify that we reproduce the same results as \citet{Salcido2016}
under their assumptions ($M_{c}=10^{10}M_{\odot}$, $p=1$, no or
short delays). In other words, their GW event rates are special cases
of our framework. As noted there, however, \citet{Sesana2007} obtained
substantially different results, as in their models seeding peaks
at $z\sim10$ and ends by $z\sim6$. In that case, $\bar{A}>0$\footnote{Specifically, $(N,\bar{A})$ for their models are: VHM: (10.67, 0.143),
KBD: (79.75, 0.26) BVRhf: (0.97, 0.076) BVRlf: (4.93,0.015).}, because the average $\log(1+z)$ is higher ($\sim1$ instead of
$\sim0.5$). Nevertheless also in those models, merger delays will
shift the GW events to lower redshifts, and thus the LISA diagnostic
diagram can diagnose the shift within a particular model. As discussed
in \citet{Salcido2016} and \citet{Barausse2012}, the number of GW
events and their redshift distribution are powerful diagnostic to
compare seeding models. The LISA diagnostic diagram, based on $(N,\bar{A})$
or just $(N,\mu)$, provides a useful visual summary of model predictions.

\section{SMBHs active as quasars at $z\sim6$}

\begin{figure}
\begin{centering}
\includegraphics[width=1\columnwidth]{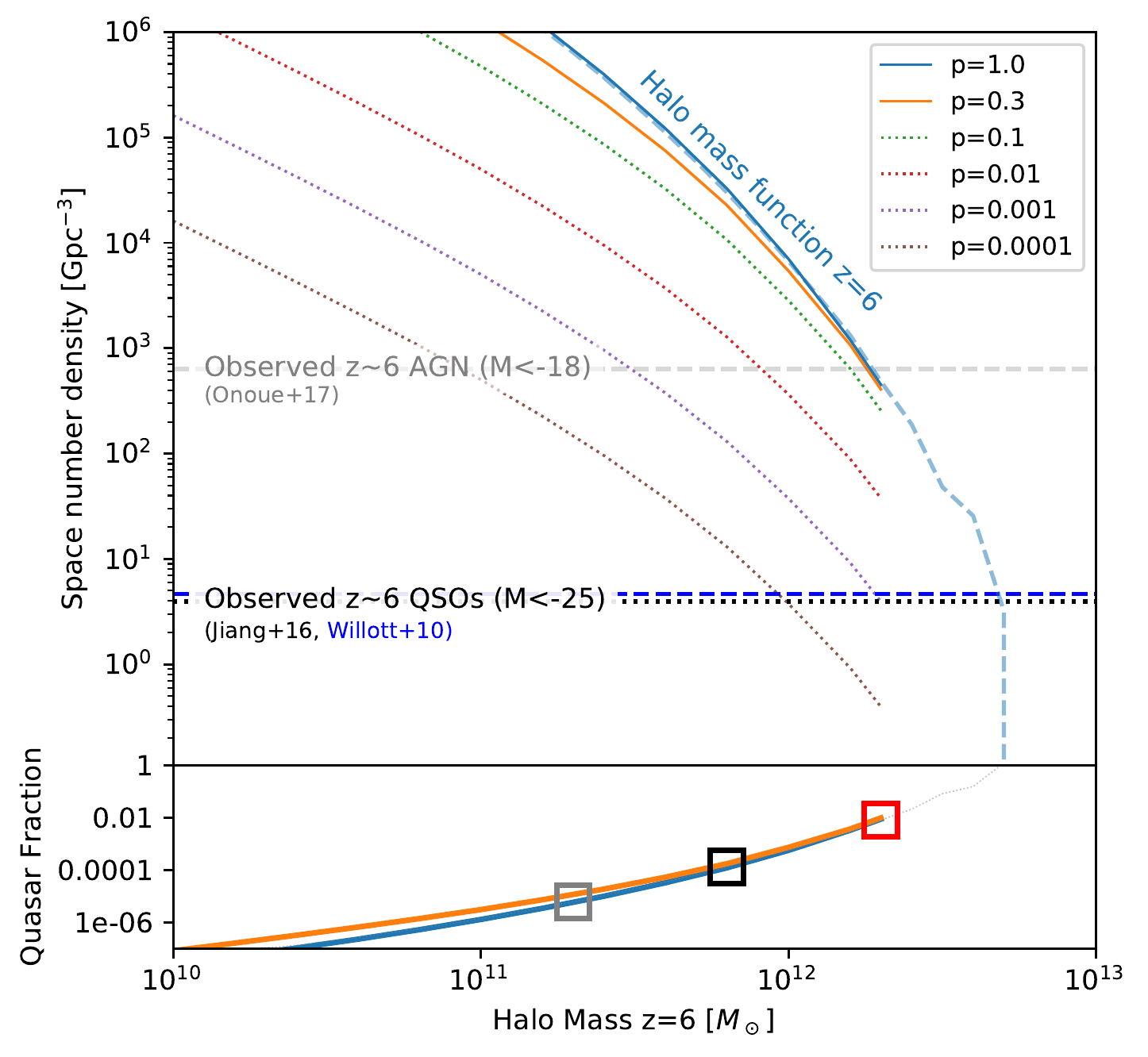}
\par\end{centering}
\caption{\label{fig:halodens6}\emph{Top panel}: Cumulative halo mass function
of black hole hosts at $z=6$. The total cumulative halo mass function
(blue curve) provides an upper envelope of the number of hosts available
above a given mass (x-axis). At very high masses this is taken from
a larger simulation box (dashed). The colored curves represent our
results from seeding halos and assuming that quasars occur in halos
above a minimum halo mass (x-axis). Surveys of bright quasars find
observed space densities of $\sim4\,\mathrm{Gpc^{-3}}$ (dashed blue
line: \citealp{Willott2010a}, dotted black line: \citealp{Jiang2016}).
Focusing on $p\protect\geq30\%$, the ratio between black hole space
density model curve and the observed quasar space density is shown
in the \emph{bottom panel}. The rectangles show three cases investigated.
Red: quasars populate only high-mass halos, black: intermediate, gray:
quasars live in $>3\times10^{11}M_{\odot}$ halos.}
\end{figure}

\label{sec:z6qsos}The earliest census of the SMBH population is available
from very high-redshift quasar surveys. The Sloan Digital Sky Survey
revealed several dozens of high-redshift optical quasars at $z\sim6-8$
\citep[e.g.,][]{Fan2001,Jiang2016}. Because quasars are interpreted
as accreting SMBHs, the quasar number density measurement places a
lower limit\footnote{It is a lower limit because this selection misses obscured, faint
and dormant black holes, and possibly, by virtue of the Eddington
limit, also those with the lowest masses. As shown before, the fraction
of dormant SMBH can be very high at least in the local Universe, and
so is the fraction of obscured AGN \citep[e.g., 75\% in][]{Buchner2015}. } on the black hole volume density. An assumption made by some previous
works is that these quasars live in the most massive halos at that
time \citep[e.g.,][]{Li2007,Volonteri2006,Wyithe2006,Romano-Diaz2011}.
This makes it easier for modelers to explain high black hole masses.
In this and the next section we explore the consequences of this halo
density matching assumption and relax it.

Figure~\ref{fig:halodens6} shows the observational lower limit as
a dotted horizontal line for quasars \citep{Willott2010a,Jiang2016}
and AGN \citep{Onoue2017}. Well above this, the top-most curve shows
the cumulative halo mass function, i.e., the number of halos above
a given mass. If quasars populated only the most massive halos, e.g.,
$M>10^{12.5}M_{\odot}$, the space density would match observations.
If instead quasars are permitted to inhabit lower masses, e.g., $M_{h}\sim10^{11}M_{\odot}$,
only a very small fraction ($f\approx10^{-6}$, bottom panel) of halos
need be quasars. According to our seeding prescription, we expect
all of these halos to have SMBHs, but their triggering as quasar requires
closer investigation.

\subsection{Environments of $z\sim6$ quasars}

\begin{figure}
\begin{centering}
\includegraphics[width=0.95\columnwidth]{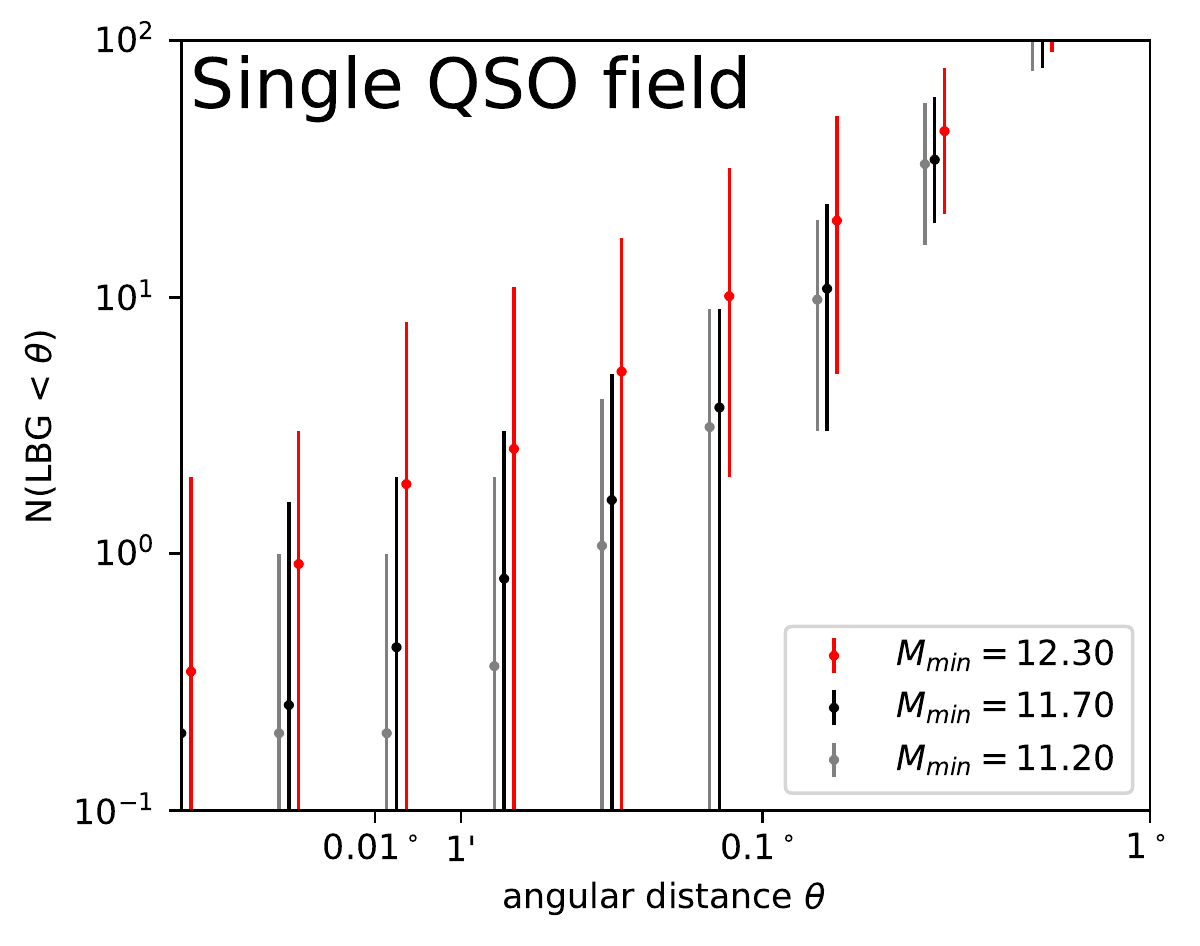}
\par\end{centering}
\begin{centering}
\includegraphics[width=0.95\columnwidth]{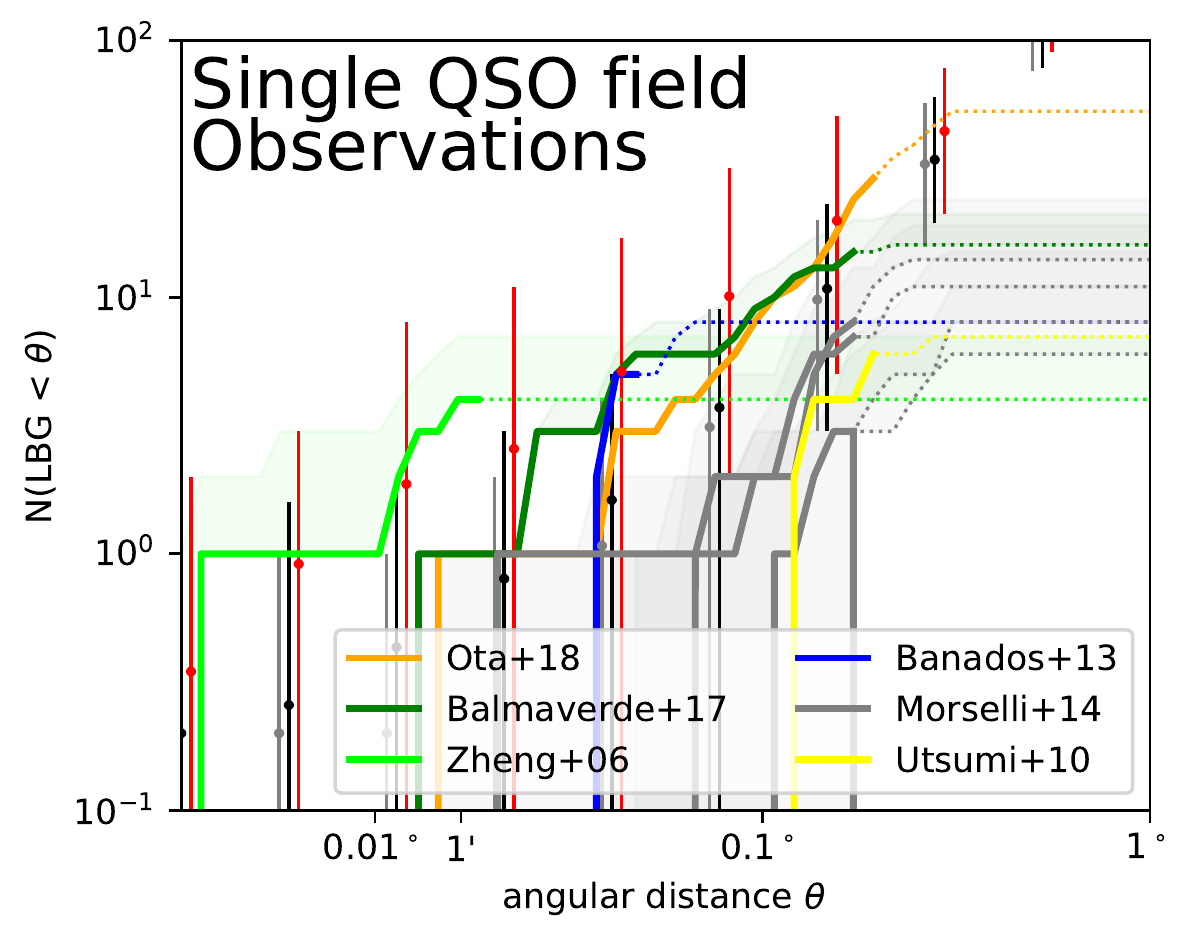}
\par\end{centering}
\begin{centering}
\includegraphics[width=0.95\columnwidth]{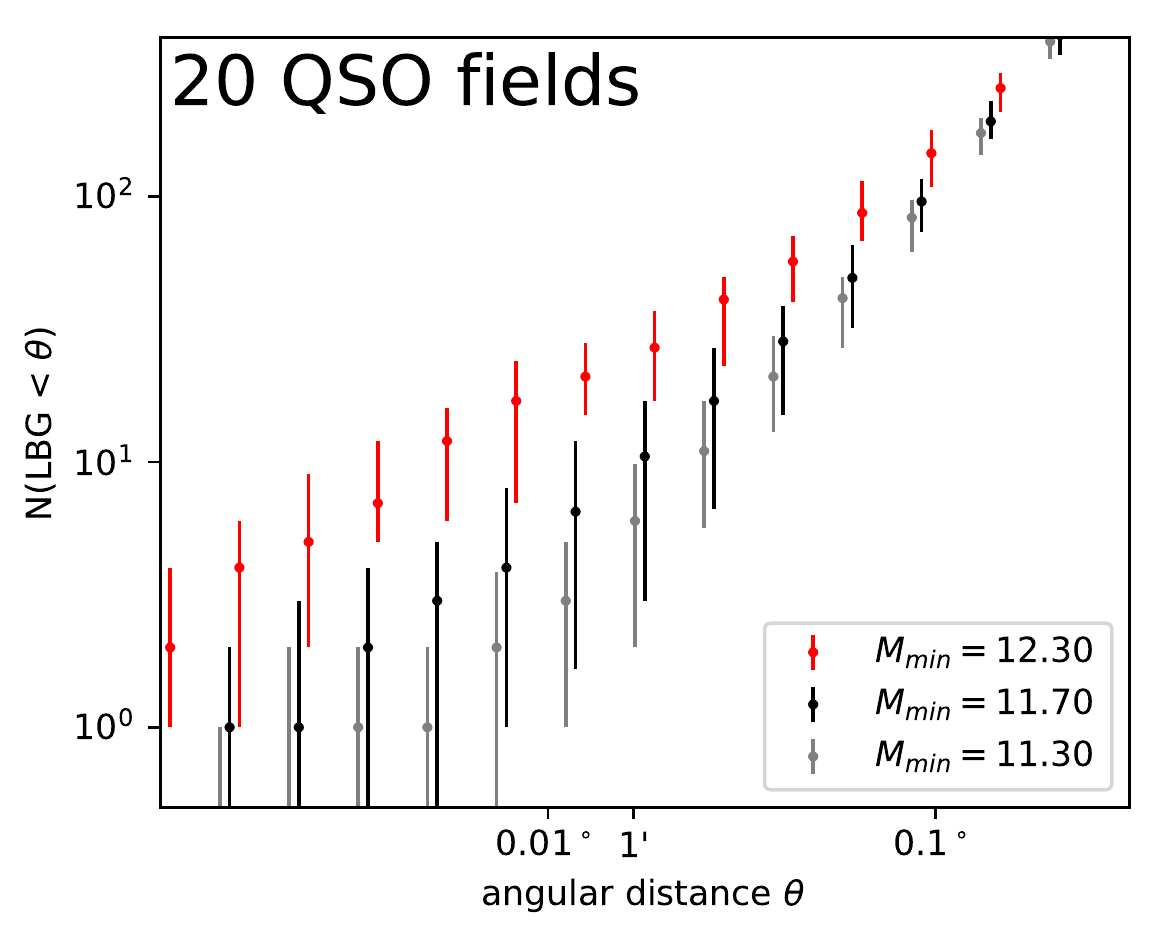}
\par\end{centering}
\caption{\label{fig:clustering-lbg}LBGs near quasars at $z\sim6$. The number
of LBGs enclosed in a given angular distance (e.g., 1~arcmin) is
plotted. \emph{Top panel}: For three cases of minimum halo masses
of quasar hosts, the frequency of LBGs is predicted. Error bars cover
95\% of randomly drawn hosts. \emph{Middle panel}: Observations. Solid
curves show the number of observed LBGs and continue as dotted when
the edge of the field is reached. When less secure candidates are
included, the numbers may be higher (shading). \emph{Bottom panel}:
If $20$ quasar fields are observed, the highest-mass model prediction
can be clearly distinguished.}
\end{figure}
\begin{figure}
\begin{centering}
\includegraphics[width=0.95\columnwidth]{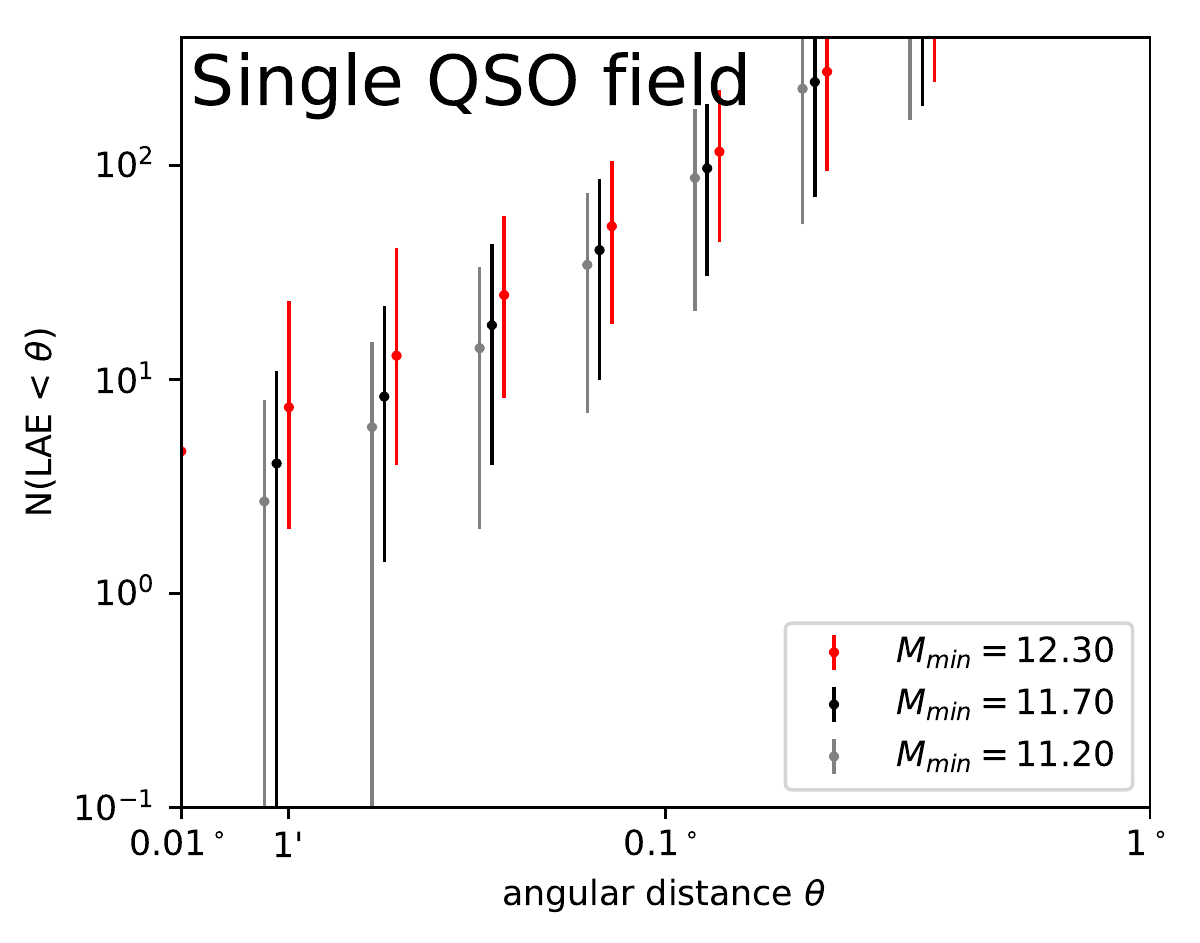}
\par\end{centering}
\begin{centering}
\includegraphics[width=0.95\columnwidth]{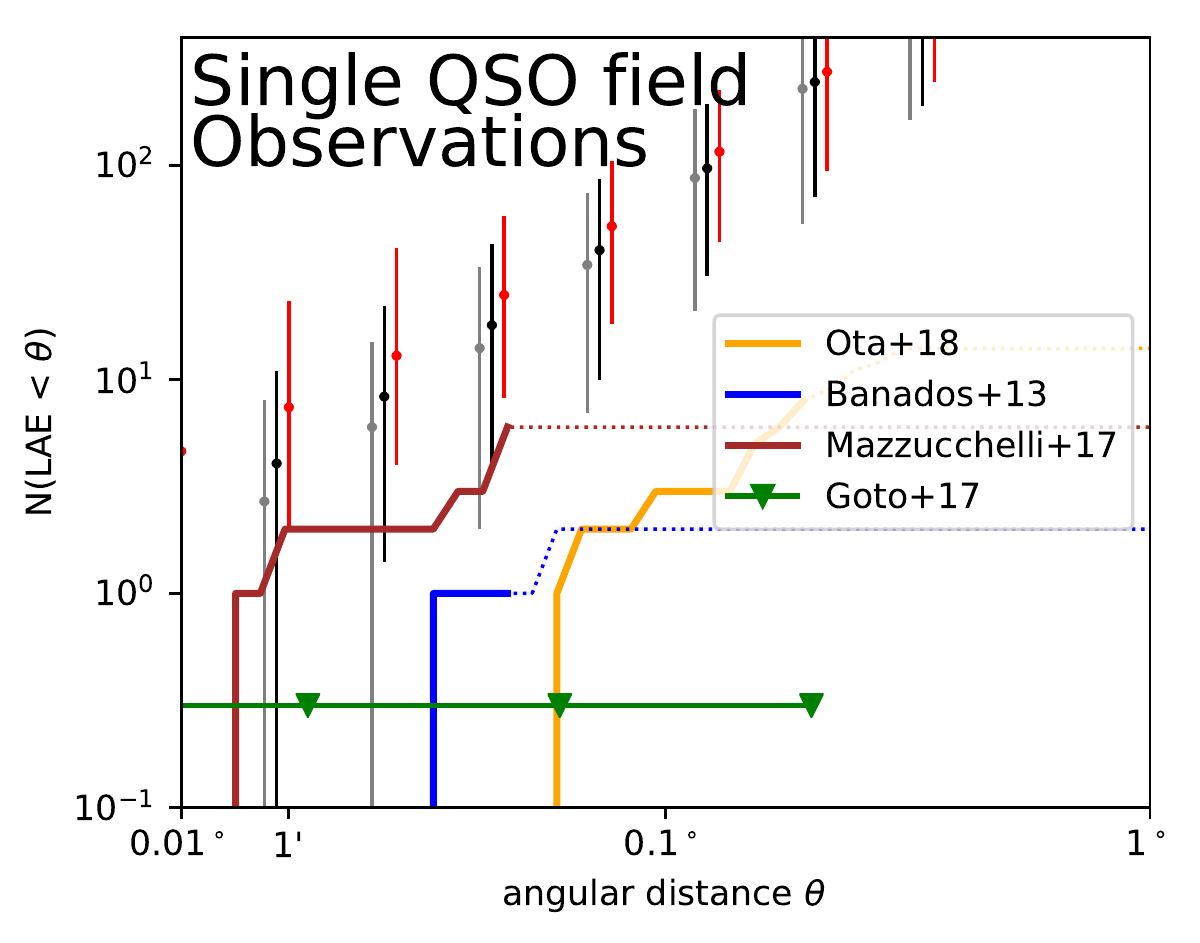}
\par\end{centering}
\begin{centering}
\includegraphics[width=0.95\columnwidth]{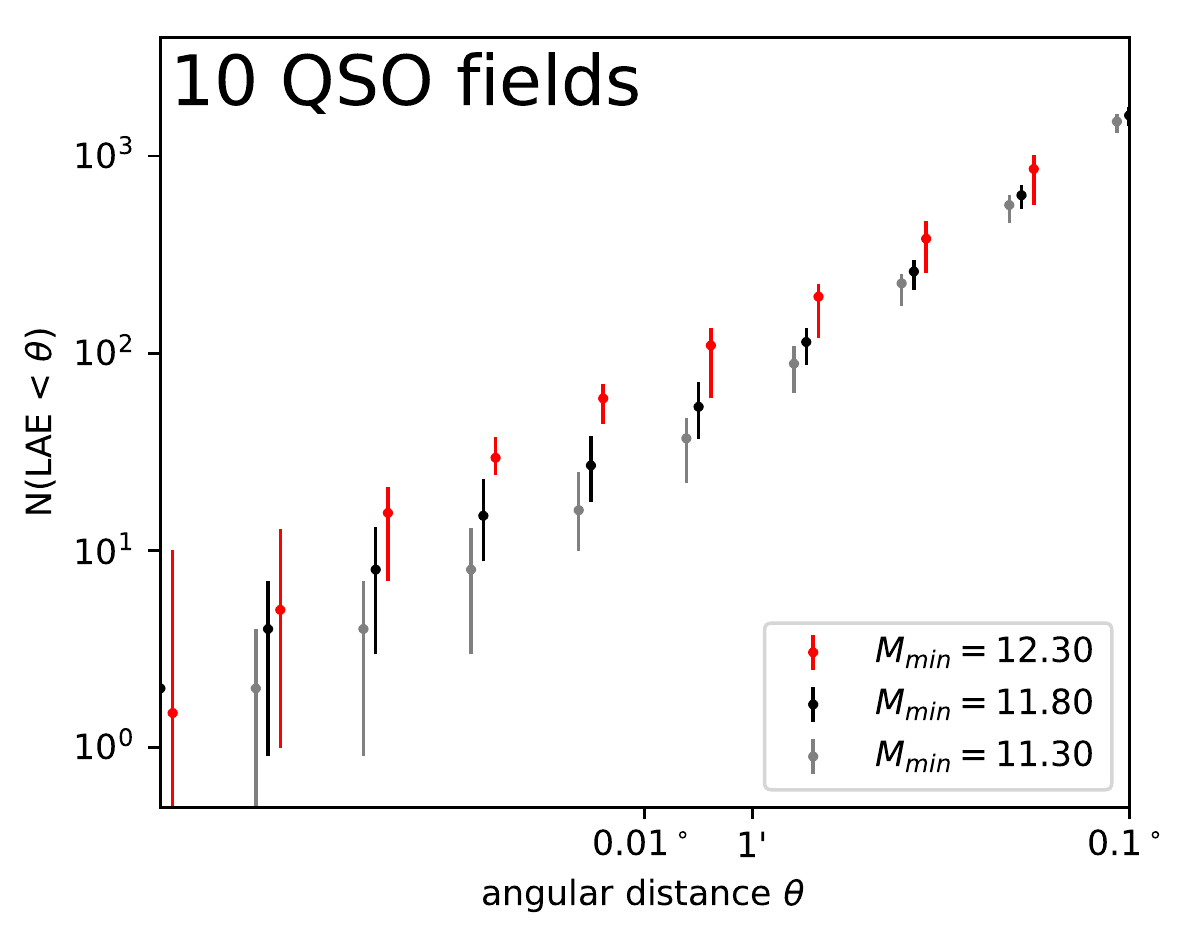}
\par\end{centering}
\caption{\label{fig:clustering-lae}LAEs near quasars at $z\sim6$. The number
of LAEs enclosed in a given angular distance (e.g., 1arcmin) is plotted.
\emph{Top panel}: For three cases of minimum halo masses of quasar
hosts, the frequency of LAEs is predicted. Error bars cover 95\% of
randomly drawn hosts. The predictions are highly degenerate. \emph{Middle
panel}: Observations. Solid curves show the number of observed LAEs
and continue as dotted when the edge of the field is reached. The
\citet{Goto2017} field (green downward-pointing triangles) did not
show any sources. \emph{Bottom panel}: If $10$ quasar fields are
observed, the model predictions start to be distinguishable.}
\end{figure}

If quasars at $z\sim6$ indeed live in very massive halos, galaxy
over-densities should be observable around them. Identifying over-densities
has however produced contradictory results in the literature. In this
section, we use mock observations to understand the difficulties.

From the simulations, we can predict the number of galaxies near quasars.
For this, we consider three cases for the minimal quasar halo masses
in the $3\times10^{11}-3\times10^{12}M_{\odot}$ range (squares in
the lower panel of Figure~\ref{fig:halodens6}), and draw 1000 halos
randomly above that threshold. For a mock observation, we then search
for surrounding galaxy halos, taking into account the angular separation
and redshift selection window. We assign galaxies to halos to mimic
observations.

Lyman break galaxies (LBGs) can be found at these redshifts through
filter drop-outs \citep[e.g.,][]{Stanway2003,Zheng2006,Bouwens2011,Ota2018}.
This selects a relatively broad redshift range (we adopt here $z=5.5-7$;
e.g., \citealp{Ota2018}) to a typical detection limit of 26th magnitude
AB \citep[e.g.,][]{Utsumi2010,Morselli2014,Ota2018}. Following the
LBG clustering analysis in the UltraVISTA and UDS fields of \citet{Hatfield2017},
we use the simple prescription that LBGs occupy halos of $M>10^{11.2}M_{\odot}$,
and verify that this matches the number detected in those (non-QSO)
fields. Selecting a cylinder as high as the redshift selection ($\sim500\mathrm{Mpc}$)
and projecting it, we take into account chance alignments. For each
quasar we compute the number of LBGs enclosed within an angular separation
of $\theta$. This is shown in the top panel of Figure~\ref{fig:clustering-lbg}.
The mean number is generally higher in the high-mass case (red), but
there is substantial overlap in the 95\% confidence intervals. This
indicates that a \emph{single quasar field is not powerful enough
to discriminate} halo mass occupations, and wide variations of numbers
are expected.

This is indeed what is observed: Some studies claim overdensities
\citep[e.g.,][]{Zheng2006}, while others find densities comparable
to control fields \citep[e.g.,][]{Banados2013,Morselli2014,Mazzucchelli2017,Ota2018}.
In the middle panel of Figure~\ref{fig:clustering-lbg} we compiled
some studies\footnote{The luminosities of the targets are consistent with the quasar definition
used in Figure~\ref{fig:halodens6}. The LBG magnitude detection
limits in these studies are comparable.} and present as thick curves the enclosed number of LBGs as a function
of angular distance. Overall, the data are consistent with all three
model ranges at all radii. The \citet{Zheng2006} data set\footnote{We have considered their C-complex as a single source and chose those
objects only $1\sigma$ above their color-cut as less secure candidates.} could be an exception just where their field-of-view ends, as well
as one of the \citet{Morselli2014} fields at the low end\nocite{Utsumi2010}.
Clearly, more observations are needed to reduce the variations. The
bottom panel of Figure~\ref{fig:clustering-lbg} shows model predictions
for 20 quasar fields. At this point the most massive scenario can
be unambiguously distinguished.

Some of the prediction variance is likely due to the broad redshift
range of the LBG selection ($\Delta z\approx1.5$ corresponding to
$\sim500\mathrm{Mpc}$). To focus on the environment near the quasar,
studies have advocated the use of Lyman Alpha emitters (LAEs) to complement
LBGs studies. To detect the Ly$\alpha$ line near the target quasar
redshift requires narrow (and sometimes custom-made) filters, but
has the benefit of probing a narrow redshift range ($\Delta z\sim0.1$).
We again adopt a simple prescription, following \citet{Kovavc2007}
and \citet{Sobacchi2015}, and assign LAEs to halos above a halo mass
limit of $M>10^{10.6}M_{\odot}$. We verify that the numbers in non-QSO
fields, SXDS \citep{Ouchi2010} and SDF \citep{Ota2018}, are reproduced.
These can also be reproduced by choosing a lower mass cut and a duty
cycle, but we find this does not change our conclusions significantly.
We again compute the enclosed number of neighbors to the quasar at
various angular separations. The top panel of Figure~\ref{fig:clustering-lae}
shows the predictions, which are indistinguishable for a single quasar
field. Observations \citep[middle panel;][]{Banados2013,Mazzucchelli2017,Goto2017,Ota2018}
fall within the predicted ranges, which are extremely wide, in part
due to Poisson statistics. The only measurement falling outside one
of the predicted ranges is that of \citet{Goto2017}, which did not
detect any sources\footnote{This is the same field as the \citet{Utsumi2010} LBG measurement.}.
This may indicate that our LAE prescription is too abundant. The bottom
panel shows predictions when 10 quasar fields have been observed.
In that case, the predictions marginally separate.

The LAE and LBG populations are both promising approaches to constrain
the halo mass of quasars, and thereby the fraction of black holes
active as quasars (Figure~\ref{fig:halodens6}). However, because
dark matter halo neighborhoods are diverse, observations of many quasar
fields are needed to make definitive statements.

\section{Discussion \& Conclusion}

We made an analysis of SMBH seeding that is independent of the seed
mass, growth mechanism and feedback processes, to estimate the number
of massive black holes across cosmic time. Our assumption, following
\citet{Menou2001}, is simply that by the time a dark matter halo
reaches a critical mass $M_{c}$, some process has seeded it with
a black hole with efficiency $p$. Details of the seeding process
involved are not needed in our analysis.

The high black hole \textbf{\emph{occupation fraction}} observed in
local low-mass galaxies constrains these parameters. The seeding fraction
$p$ has to exceed $p\apprge(M_{c}/10^{11}M_{\odot})^{3/4}$ (see
Figure~\ref{fig:fracparamspace}). This simultaneously constrains
the underlying population, for example the local SMBH space density
is above $>0.01/\mathrm{Mpc}^{3}$ in all our models. In our framework,
black hole seeding happens continuously throughout cosmic time, but
is generally most frequent at $z>4$. If we require seeding to only
occur before the epoch of reionization ($z>7$), generally lower mass
thresholds are required to increase the population (dotted line in
Figure~\ref{fig:fracparamspace}). However, the merger distributions
shift only slightly to higher redshifts. Indeed, when seeding only
a small fraction of halos (e.g., $p=1\%$), a redshift cut-off has
virtually no effect on the shape of the host merger history.

Our framework can approximate the behavior of physically-motivated
seeding mechanisms. Some examples are shown in Figure~\ref{fig:fracparamspace}
for comparison. The light seed scenario of \citet{Volonteri2003}
considers the end-products of Population III stars. This mechanism
effectively populates all $M_{c}\gtrsim10^{7}M_{\odot}$ halos at
$z\sim20$ with SMBH seeds, and as such taken as a special case of
our formalism. Heavy seed scenarios operate at higher masses $M_{c}\gtrsim3\times10^{7}M_{\odot}$
halos at $z\sim10$, but are less efficient ($p=3-30
$, e.g., \citealt{Volonteri2008,Agarwal2012}). These require UV radiation
from neighboring galaxies, which usually merge after the host halo
creates the seed \citep[by z~6, e.g.,][]{Agarwal2014}. Their effective
behavior may thus perhaps be better approximated with a higher mass
threshold with lower efficiencies down to lower redshift. In practice
however, the differences in the model predictions at lower redshift
($>1\mathrm{Gyr}$ after seeding) are negligible. If we apply a $z>15$
constraint to our seeding prescription this yields very stringent
constraints on the parameter space, ruling out $M_{c}\gtrsim10^{9}M_{\odot}$,
and giving $p\gtrsim20\%$ for $M_{c}=10^{8}M_{\odot}$, the smallest
halo masses our simulations can reliably probe. These efficiency constraints
are also consistent with \citet{Greene2012}, which showed that the
aforementioned models are near current observational constraints.
In any case however, all surviving models have populated all $M>10^{11}M_{\odot}$
halos with SMBH seeds at $z\sim6$.

\textbf{\emph{The quasar population at $z\sim6$}} can also be connected
to our prescription. \citet{Willott2010} argued that the observed
$z\sim6-7$ quasars are the $M_{\mathrm{BH}}>10^{7}M_{\odot}$ population
accreting at the Eddington-limit, while the remaining black holes
are pristine seeds without substantial accretion. Under this interpretation
we consider $f$, the success fraction of turning a seed into a quasar
at that redshift. Indeed, the above constraints indicate that at $z\sim6$,
there is an abundance of black holes. Fewer than $\sim10^{-6}$ of
the seeds are quasars at that time. If seeds become SMBH and potential
quasars only above a certain halo masses, the active fraction can
still be as low as $f\sim10^{-5}$, indicating that we may only see
very ``lucky'' seeds.

Thus, our constraints show that physical seeding mechanisms and the
activation of quasars can be highly inefficient. For example, when
studying a comoving cosmological volume of $20\,\mathrm{Mpc}$ side
length, the theorist needs to create $\sim80$ seeds by $z\approx6$,
but only one in a million of those need become a quasar with a SMBH.
It is thus encouraging to consider seeding and feeding mechanisms
requiring rare conditions, such as special configurations of halos
or multiple, complex mergers \citep[see e.g.,][]{Agarwal2012,Mayer2018,Inayoshi2018},
rather than focusing on massive halos.

Several works have studied \textbf{\emph{overdensities to measure
the halo mass of quasars}}, with mixed results. We demonstrate that
the diversity of single-QSO field observations is expected. This is
because even at a given mass, halos have diverse neighbourhoods, with
the number of surrounding halos varying by an order of magnitude.
Previously, \citet{Overzier2009} came to a similar conclusion considering
a mock field observation of i-band dropout galaxies. Observations
of several quasar fields are necessary to make definitive statements
about the host halo mass of quasars. LBGs and LAEs are both suitable
probes. However, given that detection of LAEs requires filters specific
to the targeted quasar redshift, observations of LBGs may be more
economical. In the clustering predictions we have made highly simplified
assumptions, especially in how LAEs and LBGs populate and cluster
around halos at high redshift. This is still an open research question
also in non-quasar fields.

A promising \emph{future probe} of the seeding scenarios is the space-based
LISA gravitational wave experiment \citep{eLISAConsortium2013}, because
of its broad mass and redshift sensitivity. We have presented a new
diagnostic diagram for LISA GW events. The number of observed events
and their redshift distribution can probe simultaneously both the
space density of the SMBH population and the typical delay between
halo and black hole mergers. We populate the diagram with various
seeding scenarios, taking into account possible delays between halo
mergers and black hole coalescence, and find that the local SMBH occupation
constraints already imply a lower limit of at least one GW event per
year detected by LISA. Current SMBH constraints are lower limits on
their occurance, as inactive SMBH can often go undetected. To strongly
constrain the parameter space, e.g., to near the dashed line in Figure~\ref{fig:fracparamspace},
it would be necessary to establish that 50\% of the $M_{\star}\sim10^{9}M_{\odot}$
galaxies lack a SMBH (under some mass definition of SMBH seeds). This
is challenging because with an imaging resolution $\theta$ the black
hole sphere of influence is detectable only to distances $D\approx2.3\mathrm{Mpc}\times(\theta/0.1'')\times\sqrt{M_{\mathrm{BH}}/10^{6}M_{\odot}}$
\citep[e.g.,][]{Do2014}\footnote{This contains $M_{\mathrm{BH}}\propto\sigma^{4}$, however the relation
is uncertain at low masses \citep[see, e.g.,][]{Kormendy2013,Graham2015})}. Thus many local galaxies are beyond the resolution limit of current
instruments \citep{Ferrarese2005}, but can hopefully be probed with
upcoming Extremely Large Telescopes \citep{Do2014}. Thanks to its
sensitivity to black hole mergers of various sizes, LISA will provide
the most powerful constraints on the abundance of SMBHs. Figure~\ref{fig:GW}
shows that the number of LISA detections per year correlates with
the local black hole number density, which relates to a $M_{c}\times p$
combination (right panel of Figure~\ref{fig:evolution}).

\section{Acknowledgements}

JB thanks Roberto E. Gonzalez and Nelson D. Padilla for feedback on
the manuscript. We acknowledge support from the CONICYT-Chile grants
Basal-CATA PFB-06/2007 and Basal AFB-170002 (JB, FEB, ET), FONDECYT
Regular 1141218 (FEB) and 1160999 (ET), FONDECYT Postdoctorados 3160439
(JB), CONICYT PIA Anillo ACT172033 (ET), and the Ministry of Economy,
Development, and Tourism's Millennium Science Initiative through grant
IC120009, awarded to The Millennium Institute of Astrophysics, MAS
(JB, FEB). LFS and KS acknowledge support from SNSF Grants PP00P2\_138979
and PP00P2\_166159. Discussion of some aspects of this work was carried
out at the Aspen Center for Physics, which is supported by National
Science Foundation grant PHY-1607611.

The authors gratefully acknowledge the Gauss Centre for Supercomputing
e.V. (www.gauss-centre.eu) and the Partnership for Advanced Supercomputing
in Europe (PRACE, www.prace-ri.eu) for funding the MultiDark simulation
project by providing computing time on the GCS Supercomputer SuperMUC
at Leibniz Supercomputing Centre (LRZ, www.lrz.de). The MultiDark
simulations were performed on Pleiades supercomputer at the NASA Ames
supercomputer centre. The SMDPL simulation have been performed on
SuperMUC at LRZ in Munich within the pr87yi project. The MultiDark
Database used in this paper and the web application providing online
access to it were constructed as part of the activities of the German
Astrophysical Virtual Observatory as result of a collaboration between
the Leibniz-Institute for Astrophysics Potsdam (AIP) and the Spanish
MultiDark Consolider Project CSD2009-00064. The Geryon cluster at
the Centro de Astro-Ingenieria UC was extensively used for the FORS
simulation. The Anillo ACT-86, FONDEQUIP AIC-57, and QUIMAL 130008
11 provided funding for several improvements to the Geryon cluster. 

\bibliographystyle{aasjournal}
\bibliography{all}

\end{document}